\documentclass[aps,twocolumn,prl,floatfix,superscriptaddress,showpacs,longbilbiography]{revtex4-2}
\usepackage[utf8]{inputenc} 

\usepackage{graphicx}
\usepackage{dcolumn}
\usepackage{bm}

\usepackage{braket}
\usepackage{amsfonts}
\usepackage{amsmath}
\usepackage{amssymb}

\usepackage{color,soul}
\usepackage{wasysym}
\usepackage{mathrsfs}
\usepackage{times}

\usepackage[dvipsnames,svgnames,table]{xcolor}

\usepackage{hyperref}
\usepackage{fancyhdr}
\usepackage{float}

\usepackage[english]{babel}

\usepackage[caption=false]{subfig}
\usepackage{braket}

\hypersetup{
pdfstartview={FitH},
colorlinks=true,    
linkcolor=NavyBlue, 
citecolor=Maroon,   
filecolor=NavyBlue, 
urlcolor=NavyBlue   
}

\newcommand{\bea}{\begin{eqnarray}}
\newcommand{\eea}{\end{eqnarray}}

\begin{document}

\title{Copropagating edge states produced by the interaction between electrons and chiral phonons in two-dimensional materials}

\author{Joaqu\'{\i}n {Medina Due\~nas}}
\affiliation{Departamento de F\'{\i}sica, Facultad de Ciencias F\'{\i}sicas y Matem\'aticas, Universidad de Chile, Santiago, Chile}
\author{Hern\'an L. Calvo}
\affiliation{Instituto de F\'{\i}sica Enrique Gaviola (CONICET) and FaMAF, Universidad Nacional de C\'ordoba, 5000 C\'ordoba, Argentina}
\author{Luis E. F. {Foa Torres}}
\affiliation{Departamento de F\'{\i}sica, Facultad de Ciencias F\'{\i}sicas y Matem\'aticas, Universidad de Chile, Santiago, Chile}

\begin{abstract}
Unlike the chirality of electrons, the intrinsic chirality of phonons has only surfaced in recent years. 
Here we report on the effects of the interaction between electrons and chiral phonons in two-dimensional materials by using a non-perturbative solution. We show that chiral phonons introduce inelastic \textit{Umklapp} processes resulting in copropagating edge states which coexist with a continuum. Transport simulations further reveal the robustness of the edge states. Our results hint on the possibility of having a metal embedded with hybrid electron-phonon states of matter.

\end{abstract}

\date{\today}
\maketitle

\textit{Introduction.--} Since the very beginning of the quantum theory of solids~\cite{bridgman_electrical_1921,peierls_zur_1930}, the interaction between electrons and lattice vibrations has provided a long list of exciting discoveries and its effects have proven to be ubiquitous in condensed matter physics. Hallmarks of this interaction pervade in three dimensional materials~\cite{ziman_electrons_2001}, as well as in low dimensional systems~\cite{challis_electron-phonon_2003,samsonidze_electron-phonon_2007,yan_electron-phonon_2013,gunst_first-principles_2016,ersfeld_spin_2019}. Prominent examples include the role played by electron-phonon (e-ph) interaction in the development of the theory of superconductivity~\cite{frohlich_theory_1950,bardeen_electron-vibration_1951,tinkham_introduction_2004} and conducting polymers~\cite{heeger_solitons_1988}, where charge doping is used to circumvent the Peierls transition~\cite{peierls_quantum_1955,peierls_more_1991}. More recently, different studies have pointed the possibility of electron-phonon-induced bandgaps~\cite{foa_torres_inelastic_2006,foa_torres_nonequilibrium_2008}, robust edge states~\cite{calvo_robust_2018}, and phonon-induced topological phases~\cite{hubener_phonon_2018,chaudhary_phonon-induced_2020}.

In 2015 a new twist in this field was triggered by the prediction of phonons with \textit{intrinsic} chirality in monolayer materials~\cite{zhang_chiral_2015}. Further theoretical studies~\cite{liu_pseudospins_2017,gao_nondegenerate_2018,xu_nondegenerate_2018,chen_chiral_2019} and the experimental observation of circular phonons in monolayer tungsten diselenide~\cite{zhu_observation_2018} brought a focus to this overlooked property. Thanks to the breaking of inversion symmetry, the degeneracy between clockwise and counterclockwise modes can be lifted at high symmetry points of the Brillouin zone (BZ), leading to the intrinsic chirality of the modes. Today, chiral phonons are at a focal point in the context of the pseudogap of the cuprates~\cite{grissonnanche_chiral_2020,torre_mirror_2021}, where phonons have been proposed to become chiral in the pseudogap phase of these materials~\cite{grissonnanche_chiral_2020}. Recent experiments have shown electrostatic control over the angular momentum of phonons~\cite{sonntag_electrical_2021}. A proposal for chiral phonons with  non-vanishing group velocity~\cite{chen_propagating_2021} also adds interest to this blooming area. 
But, notably, the effects of e-ph interaction with chiral phonons remains mostly unexplored.
Therefore, it is of interest to address this issue. One may wonder whether the chirality of phonons could give rise to any unusual effects on the electronic structure through their mutual interaction, or whether new hybrid e-ph states could be controlled by tuning either the electronic or phononic degrees of freedom. 

\begin{figure}[tb!]
\centering
\includegraphics[width=0.8\linewidth]{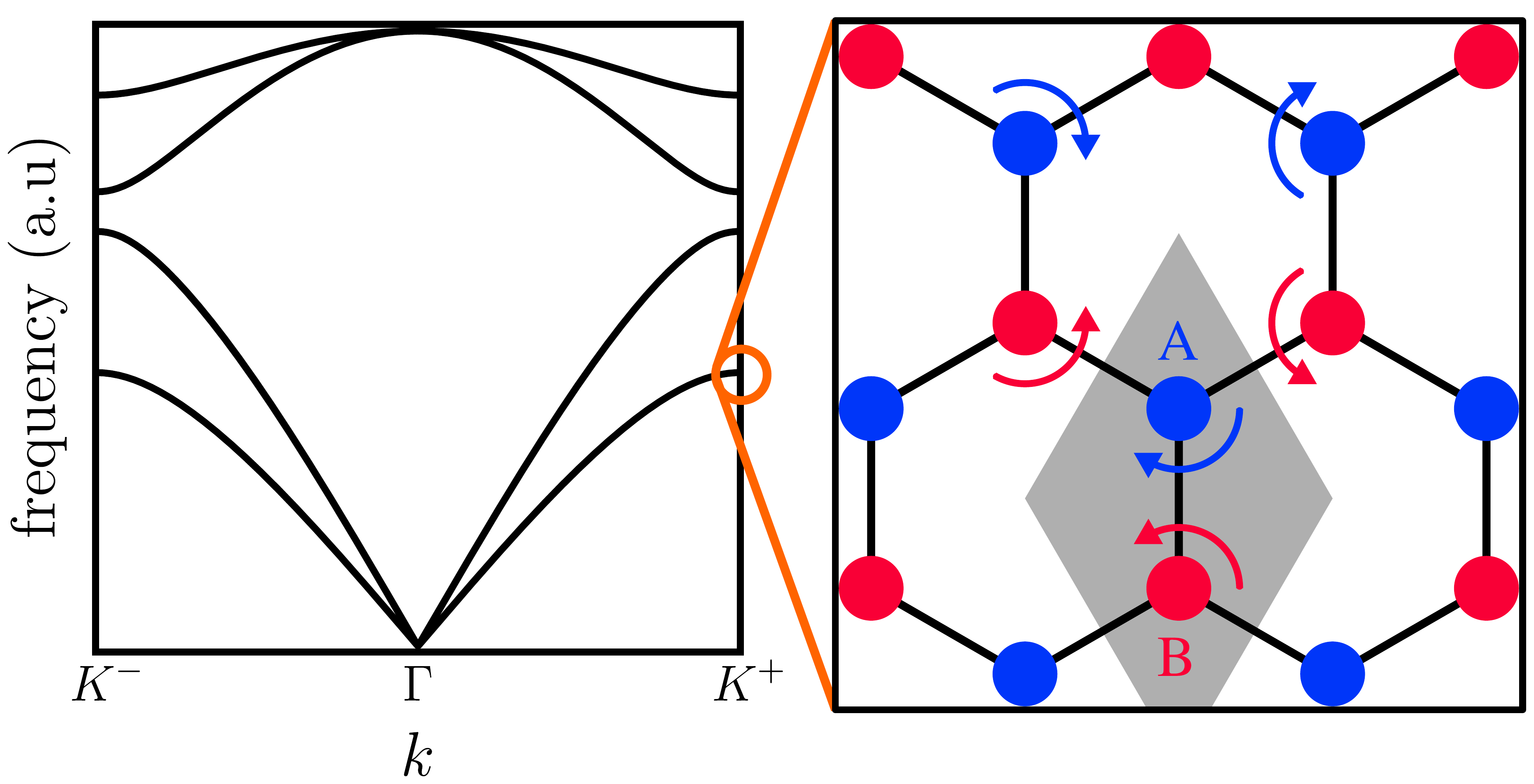}
\caption{Scheme depicting the chiral phonon mode considered during this work. The left panel shows the phonon band structure of a honeycomb lattice, in arbitrary units, with chiral phonons at high symmetry points. Breaking inversion symmetry lifts the degeneracy between the two middle bands at the valleys, and generates circular motion of the sublattices with different amplitudes, allowing for phonon modes with net chirality~\cite{zhang_chiral_2015}. We focus on the mode represented in the right panel, where sublattice A (B), depicted by blue (red) circles, exhibits counter-clockwise (clockwise) circular motion, as indicated by the corresponding arrows. Due to the non-zero phonon momentum, the movement acquires a phase $e^{\pm i2\pi/3}$ between neighboring unit cells, represented by the arrows.}
\label{fig:fig1}
\end{figure}

Here we explore the effects of e-ph interaction with chiral phonons in two-dimensional materials with broken inversion symmetry, looking for novel hybrid e-ph states of matter. Aided by a non-perturbative and non-adiabatic Fock space solution, we explore the effect of the interaction with a single chiral phonon mode as depicted in Fig.~\ref{fig:fig1}. A first finding is that the chirality of the phonons gets imprinted in the e-ph interaction term, which breaks time-reversal symmetry while introducing inelastic \textit{Umklapp} processes (which are not present, for example, in the case of laser-illuminated materials where the transitions are vertical~\cite{rudner_band_2020}). We show that this interaction opens a gap in one of the two inequivalent valleys which is bridged by two hybrid e-ph edge states. Interestingly, these edge states turn out to propagate in the \textit{same} direction on opposite edges and coexist with a continuum of states in the ungapped valley. Further quantum transport simulations demonstrate that the copropagating edge states are robust, as they can withstand even moderate amounts of short-range disorder. As a playground for realizing this physics, we put forward the case of graphene on hexagonal boron nitride (hBN) for which we provide more details.

\textit{Hamiltonian model.---}
We consider a honeycomb lattice including the interaction between electrons and a single phonon mode. The Hamiltonian is written as: ${\cal H} = {\cal H}_\text{e} + {\cal H}_\text{ph} + {\cal H}_\text{e-ph}$,
where ${\cal H}_\text{e}$ and ${\cal H}_\text{ph}$ are the independent electronic and phonon contributions, while ${\cal H}_\text{e-ph}$ is the interaction term. The electronic contribution is modeled as a simple nearest neighbors Hamiltonian for a honeycomb lattice with a single orbital per atom and a staggered onsite term accounting for the inversion symmetry breaking:
\begin{equation}
    {\cal H}_\text{e} = \sum_n \Delta_{n} c_n^\dag c_n + \gamma_0 \sum_{\left\langle n, m \right\rangle} c_n^\dag c_m \text{ ,}
\end{equation}
where $c_n$ stands for the electronic annihilation operator at site $n$. The second sum on the right hand side runs over nearest neighbors, $\gamma_0$ is the nearest-neighbors hopping parameter. The on-site energy $\Delta_{n}$ models a staggering potential and is equal to $\Delta$ or $-\Delta$ if the site belongs to sublattice A or B. For the phonons we consider a single mode, which we choose as a chiral mode with momentum $\mathbf{G}$ and frequency $\omega$, described by: ${\cal H}_\text{ph} = \hbar\omega a^\dag a$,
with $a$ the phonon annihilation operator.

The e-ph interaction term arises from the change in the interatomic distance produced by the phonons, this naturally leads to a Su-Schrieffer-Heeger form~\cite{su_solitons_1979}, which contrasts for example with Holstein's model that typically applies to narrow band systems (see~\cite{ortmann_theory_2009} and references therein). It follows from quantizing the linear correction to the hopping amplitudes due to the atomic displacements from equilibrium.   
The displacement of site $n$ respect to its equilibrium position is $\delta\mathbf{r}_n(t) = \text{Re} \left[A e^{-i(\omega t - \mathbf{G}\cdot\mathbf{R}_n)} \mathbf{u}_\nu \right]$,
with $\mathbf{R}_n$ the lattice vector of the corresponding unit cell, $A$ the amplitude of the motion, $\nu$ indicating the sublattice of site $n$ ($\nu = \text{A, B}$), and $\mathbf{u}_\nu$ the eigenvector of the phonon mode on sublattice $\nu$. 
Phonon chirality is given by a circular motion of the sites, represented in the phonon eigenvector $\mathbf{u}$, which constitutes an intra-cell contribution to the chirality, as well as by an inter-cell term given by the phase $e^{i\mathbf{G}\cdot\mathbf{R}_n}$ acquired by the motion in the different unit cells~\cite{zhang_chiral_2015}.
We incorporate the lattice vibrations to the tight-binding Hamiltonian as a renormalization of the electronic hopping amplitude between sites $n$ and $m$~\cite{calvo_robust_2018}, 
\begin{equation}
    \gamma_{n,m} = \gamma_0 \exp{\left[ -b \left( \frac{|\mathbf{r}_n - \mathbf{r}_m|}{a_0} - 1 \right) \right]} \text{ ,}
    \label{eq:hoppingrenormalization}
\end{equation}
with $a_0$ the equilibrium nearest neighbor distance, $b$ the decay rate, and $\mathbf{r}_n = \mathbf{r}_n^0 + \delta\mathbf{r}_n$, with $\mathbf{r}_n^0$ the equilibrium position of site $n$. We assume small vibration amplitudes, $|A| \ll a_0$, where the zero-th order term resulting from Eq.~(\ref{eq:hoppingrenormalization}) accounts for the bare hoppings represented in ${\cal H}_\text{e}$. The first order terms couple electronic degrees of freedom with lattice vibrations, obtaining the e-ph hopping amplitudes
\begin{equation}
    \gamma_{n,m} = - \gamma_1 \hat{\mathbf{r}}_{n,m} \cdot \left[ \left( e^{i\mathbf{G}\cdot\mathbf{R}_n} \mathbf{u}_\nu - e^{i\mathbf{G}\cdot\mathbf{R}_m} \mathbf{u}_\mu \right) e^{-i\omega t} + \text{c.c.} \right] \text{ ,}
    \label{eq:e-phhopping}
\end{equation}
where $\gamma_1 \ll \gamma_0$ sets the strength of e-ph interactions, $\hat{\mathbf{r}}_{n,m} = \left(\mathbf{r}_n^0 - \mathbf{r}_m^0 \right) / a_0$, and $\nu$ ($\mu$) indicates the sublattice of site $n$ ($m$). After quantizing the lattice vibrations and Fourier transforming (for details see~\cite{supp-info}), the interaction Hamiltonian reads
\begin{equation}
\begin{split}
    \mathcal{H}_\text{e-ph} = & -\gamma_1 a^\dag \Big[ \int_\text{BZ} \text{d}^2\mathbf{k} \sum_{\nu\neq\mu} c_{\nu,\mathbf{k}}^\dag c_{\mu,\mathbf{k}+\mathbf{G}} \sum_m e^{-i(\mathbf{k}+\mathbf{G})\cdot\mathbf{R}_{n,m}} \\ & \hat{\mathbf{r}}_{n,m} \cdot \left( \mathbf{u}_\nu - e^{-i\mathbf{G}\cdot\mathbf{R}_{n,m}} \mathbf{u}_\mu \right) \Big] + \text{h.c.} \text{ ,}
    \label{eq:e-phhamiltonian}
\end{split}
\end{equation}
where the first sum runs over the sublattices, and the second sum runs over all sites $m$ of sublattice $\mu$, which are nearest neighbors of a given site $n$ of sublattice $\nu$. The term explicitly written in Eq.~\eqref{eq:e-phhamiltonian} accounts for momentum conserving phonon emission processes, 
where an electron with momentum $\mathbf{k}+\mathbf{G}$ is annihilated, while an electron with momentum $\mathbf{k}$ and a phonon with momentum $\mathbf{G}$ are created.

\begin{figure}[pbt!]
\centering
\includegraphics[width=\linewidth]{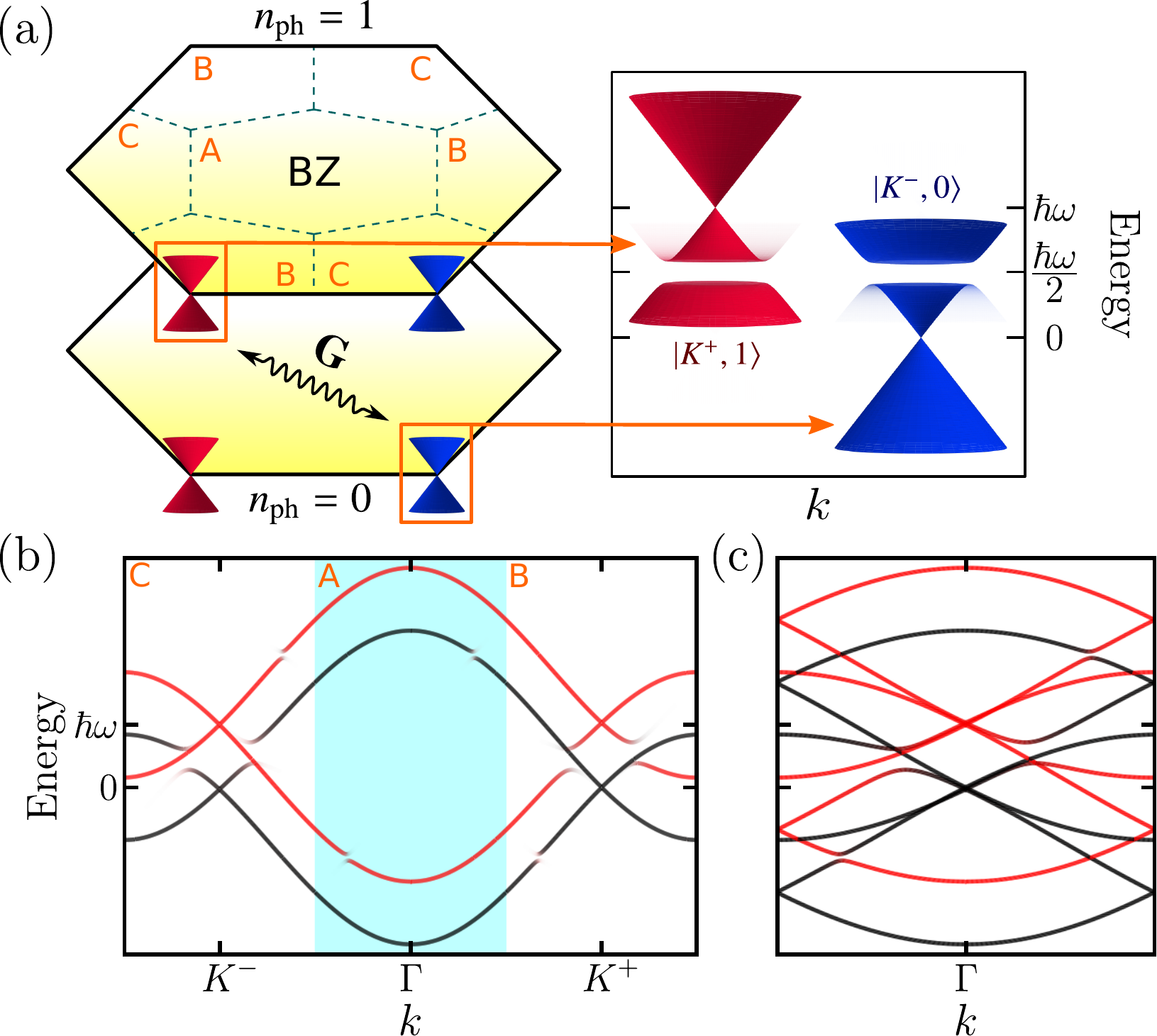}
\caption{\textbf{(a)} Considering the phonon Fock space, the system can be viewed as a semi-infinite series of pure electronic Hamiltonian replicas centered at energies $n_\text{ph}\hbar\omega$, with $n_\text{ph} \in \mathbb{N}_0$ the corresponding phonon population. e-ph interactions generate non-vertical transitions in the BZ, between momentum $\mathbf{k}$ in replica $n_\text{ph}=0$ to momentum $\mathbf{k}-\mathbf{G}$ in replica $n_\text{ph}=1$, with $\mathbf{G}=K^+$ the phonon momentum. A pseudogap opens in the non-vertical intersection between the Dirac cones at valleys $K^-$ and $K^+$ in replicas $n_\text{ph}=0$ and $n_\text{ph}=1$ respectively. \textbf{(b)} Band structure of the bulk system where the valley selective gap is seen at the replica crossing. The color scale indicates the weight of the bands on $n_\text{ph}=0$ (black) and $n_\text{ph}=1$ (red). \textbf{(c)} A zone-folding scheme may be developed, where areas B and C of the BZ of the hexagonal lattice are folded into area A [as indicated in panels (a) and (b)], forming the rBZ where the system presents vertical transitions. The band structure in the rBZ is represented in panel (c), where the $k$-path plotted corresponds to that of the cyan shaded area in panel (b).}
\label{fig:fig2}
\end{figure}

In what follows we consider the phonon momentum $\mathbf{G}$ corresponding to valley $K^+$, where atoms of sublattice A and B exhibit clockwise and counter-clockwise motion respectively, as depicted in Fig.~\ref{fig:fig1}. 
The motion of the sites acquires a phase $e^{\pm i2\pi/3}$ between neighboring unit cells, thus, the non-zero phonon momentum effectively modifies the periodicity of the lattice. We may however retain the original unit cell and in this case the phonon momentum generates non-vertical transitions in the original BZ.
A zone-folding scheme may be developed in this case so the system presents vertical transitions in a reduced BZ (rBZ), folding areas B and C of the BZ into area A, as depicted in Fig.~\ref{fig:fig2}, taking advantage of the threefold periodicity of the inter-cell phonon term ~\cite{supp-info}.

\begin{figure}[tbp!]
\centering
\includegraphics[width=\linewidth]{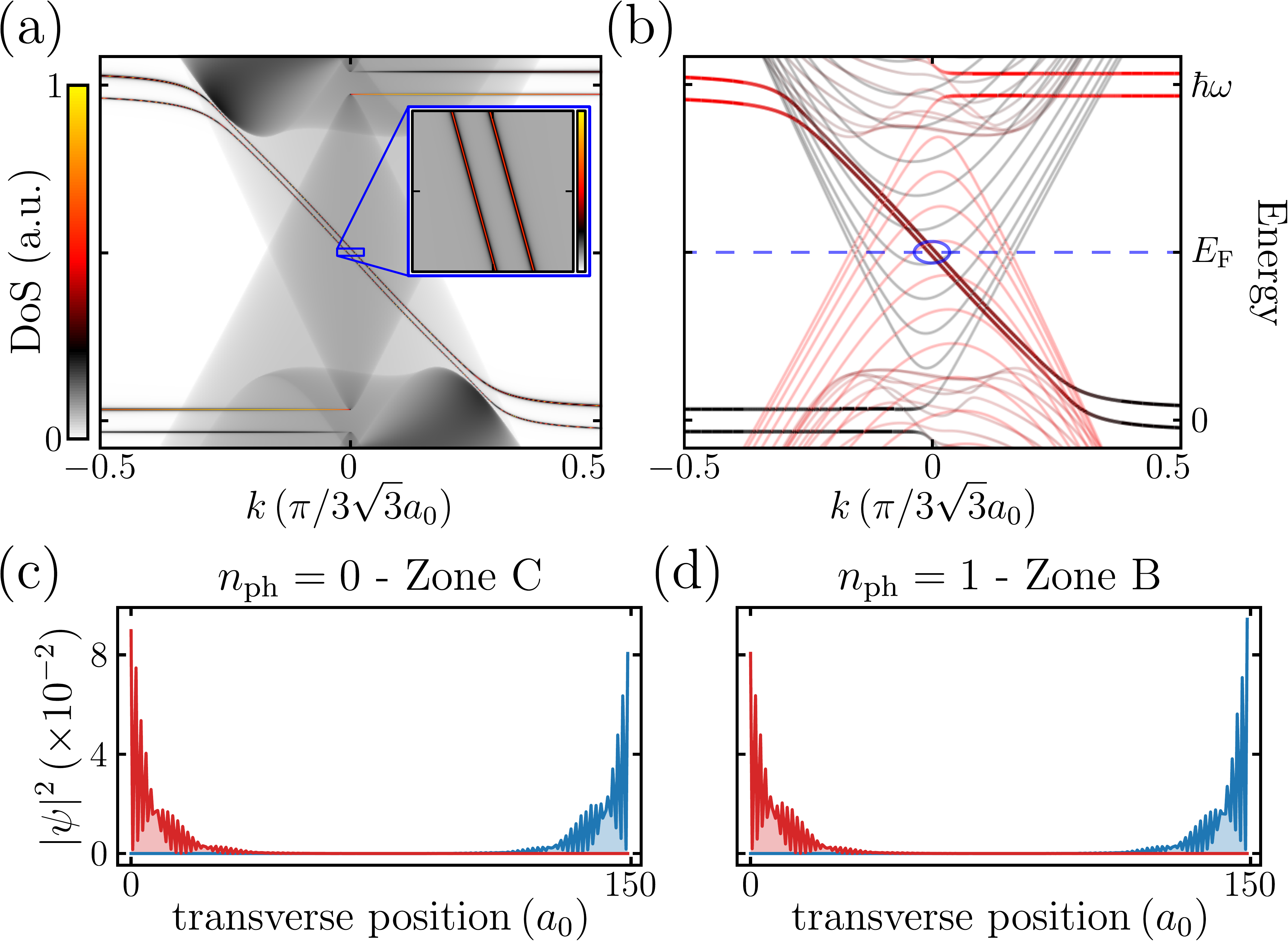}
\caption{\textbf{(a)} Density of States (DoS), shown in logarithmic scale $\text{ln}(1 + \text{DoS})$, near the edges of a semi-infinite zigzag ribbon, in arbitrary units. A valley selective gap opens in the replica crossing, bridged by two edge states with parallel velocities. These states coexist with a continuum of extended states of the ungapped valleys. \textbf{(b)} Band structure of a zigzag ribbon (of width $150 a_0$). The color scale indicates the weight of the bands on $n_\text{ph}=0$ (black) and $n_\text{ph}=1$ (red), 
while the curves' transparency is related to the localization of the states (ranging from light curves for extended states, to opaque curves for states with an inverse participation ratio larger than $0.1$)
The edge states described in (a), at energy $E_\text{F}=\hbar\omega/2$, are indicated with a blue circle. The horizontal axis in (a) and (b) corresponds to the rBZ projected to the periodicity direction. \textbf{(c)} and \textbf{(d)} Projection of the edge states in (b) to zone C of the BZ in  $n_\text{ph}=0$, and zone B in $n_\text{ph}=1$, showing that they are localized at opposite borders of the ribbon.
Other parameters used are $\gamma_1=0.025\gamma_0$, $\hbar\omega=0.3\gamma_0$, and $\Delta=0.01\gamma_0$.}
\label{fig:fig3}
\end{figure}

\textit{Band structure: Gaps and edge states.---} 
Instead of treating the e-ph interaction perturbatively as it is most usual, here we will use the non-perturbative and non-adiabatic approach introduced in ~\cite{anda_electron_1994,bonca_effect_1995}. The main idea is the exact mapping of the many body problem onto a one-particle problem in a higher-dimensional space, where each phonon mode introduces an additional dimension to the original electronic problem. This can be visualized after writing the  problem in a Fock space basis: The full Hamiltonian can be viewed as a semi-infinite series of \textit{replicas} of the original purely electronic problem centered at energies $n_\text{ph}\hbar\omega$, with $n_\text{ph}\in\mathbb{N}_0$, coupled by the interaction Hamiltonian. The phonon induced non-vertical transitions couple valley $K^-$ in $n_\text{ph}=0$ with valley $K^+$ in $n_\text{ph}=1$, generating an indirect valley selective gap at the replica crossing, as shown in Fig.~\ref{fig:fig2}-(b). The other valleys, namely $K^+$ in $n_\text{ph}=0$ and $K^-$ in $n_\text{ph}=1$, are not connected by the phonon mode and are therefore not gapped. As a consequence, the system does not present a global gap at the replica crossing, but rather a valley selective pseudogap. For $\hbar\omega \ll \gamma_0$ the valleys may be described by massive Dirac Hamiltonians. Within this regime we estimate the magnitude of the pseudogap, which is $3\sqrt{2}\gamma_1 | \mathbf{L}^\dag \mathbf{u}_\text{A} + \mathbf{R}^\dag \mathbf{u}_\text{B}|$, with $\mathbf{L} = (1,\, -i)^\text{T} / \sqrt{2}$ ($\mathbf{R} = (1,\, i)^\text{T} / \sqrt{2}$) representing the counter-clockwise (clockwise) circular motion of the atoms. Interestingly, the magnitude of the gap is sensitive to the chirality of each sublattice.

For a ribbon geometry, we expect to observe the effects of e-ph interaction in any geometry that allows to distinguish processes occurring at the distinct valleys. We focus on the energy dispersion projected on replicas $n_\text{ph}=0$ and $n_\text{ph}=1$. For a zigzag ribbon we observe two edge states in the replica crossing energy region, which correspond to a hybridization of the flat bands of each replica, as shown in Fig.~\ref{fig:fig3}. Projection of the states to the different momentum and phonon subspaces, as shown in Fig.~\ref{fig:fig3}-(c) and (d) proves that they correspond to localized edge states which bridge the valley selective bulk band gap. 
Remarkably, we find that the edge states propagate in the same direction at the opposite borders of the ribbon, as we further clarify in the next section. The remaining states that coexist at the same energy $\hbar\omega/2$ do not hybridize with their replicas, and correspond to the continuum of bulk states from the ungapped valleys.

\begin{figure*}[tbp!]
\centering
\includegraphics[width=\textwidth]{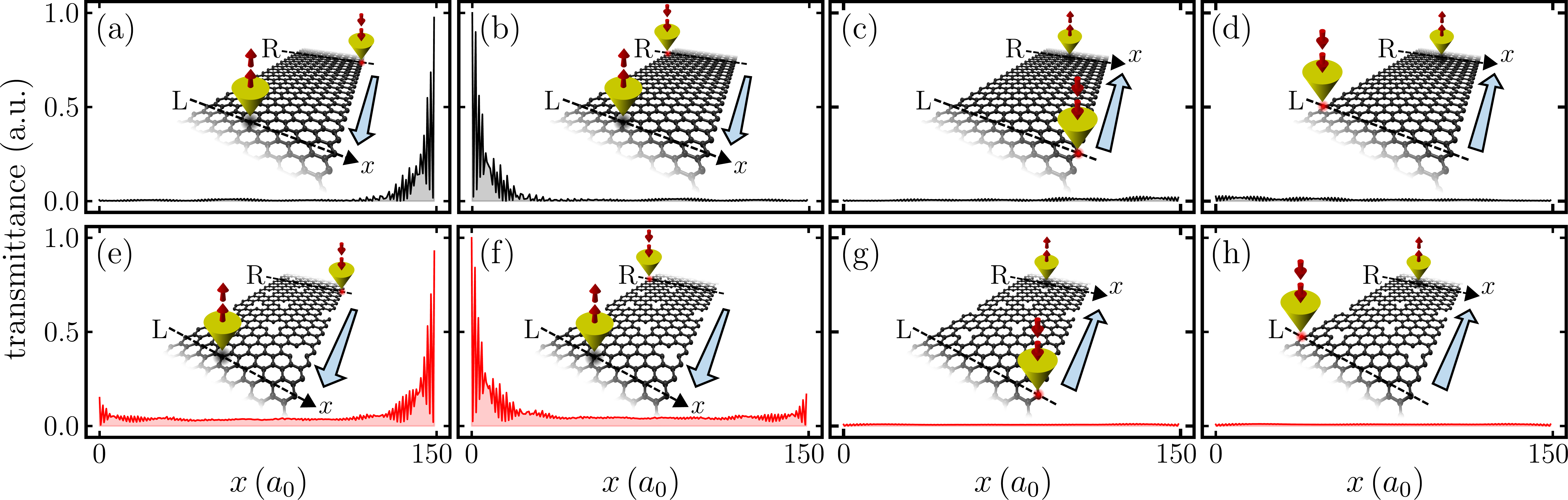}
\caption{Transmittance, in arbitrary units, 
between non-invasive probes located on different transverse lines across the zigzag ribbon (of width $150a_0$) separated by a distance $300 \sqrt{3} a_0$ (the rest of the parameters are taken as in Fig.~\ref{fig:fig3}). The input site is taken at the edge of the corresponding line, while the output site runs over the entire output line, with transverse position $x$. For easier visualization, the situation for each panel is represented as an inset, where the input sites are marked in red and signaled by downwards red arrows. The transmittance includes both the elastic and inelastic contributions and the energy of the incoming electrons is set to $\hbar\omega/2$. Panels on the top row [(a)--(d)] correspond to the pristine system while those on the lower row [(e)--(h)] include disorder (random vacancies at 0.1\%), where each of set of plots is normalized separately to its maximum value. 
}
\label{fig:fig4}
\end{figure*}

For ribbons with Klein edges we find edge states of the same nature, which bridge the valley selective gap and coexist with a continuum of extended states, though their propagation direction is reversed with respect to the zigzag case ~\cite{supp-info}. We interpret this in terms of the flat bands in honeycomb lattices with different terminations: ribbons with Klein edges develop flat bands in a $k$-space region complementary to those with zigzag edges~\cite{yao_edge_2009}. A different scenario is presented for an armchair geometry, where processes occurring at both valleys are not distinguishable, thus the consequences of the valley selective bulk gap are not observed~\cite{supp-info}. Finally, we mention other related works where copropagating edge states were proposed based on a single particle picture, either as a modified Haldane term~\cite{colomes_antichiral_2018} or in twisted graphene multilayers~\cite{denner_antichiral_2020}.

\textit{Quantum transport fingerprints of copropagating e-ph edge states.---} One may wonder about possible hallmarks of the electron-chiral-phonon interaction that might be imprinted on the transport properties. To address this question we consider a zigzag ribbon with e-ph interaction in the entire system, and compute the total transmittance (including elastic and inelastic processes) between two local probes weakly coupled to a single site located along the lines L and R of the ribbon, as depicted in the insets of Fig.~\ref{fig:fig4}. 
We choose the input site at each edge of one line (marked with a red dot in Fig.~\ref{fig:fig4}) and compute the transmittance to a site on the other line with transverse position $x$ (see ~\cite{supp-info} for further details).

Fig.~\ref{fig:fig4}-(a)--(d) shows the transmittance for a pristine ribbon, while panels (e)--(h) show the results including random vacancies with a density of $0.1\%$, averaged over 100 random realizations. A first feature revealed by our simulations is a strong directional asymmetry, transport is strongly suppressed in one direction (from L to R as shown in panels (c) and (d)) 
irrespective of the source being at one edge or the opposite, while in the reverse direction it is strongly focused near the same edge as the source 
(as shown in panels (a) and (b)). The small non-vanishing transmittance in panels (c) and (d) is attributed to a contribution from bulk states. 
This behavior is consistent with the nature of the copropagating edge states signaled earlier close to $\hbar\omega/2$. 

Panels (e)--(h) show that the behavior obtained for the pristine system withstands a moderate amount of disorder.  Disorder generates edge to edge scattering in the R to L direction, represented by an increased bulk contribution and a small peak at the opposite edge of the input site. However, the spatial distribution remains strongly peaked at the input edge. 
Importantly as well, the directional asymmetry in the transmittance persists entirely, even against this worst-case scenario of short range disorder.

\textit{Final remarks.---} Our results show a first glimpse of the effects of electron-chiral-phonon-interaction in two-dimensional materials. The interaction with chiral phonons provides for a novel effective time-reversal symmetry breaking term. Unlike other known symmetry breaking terms, such as a magnetic field~\cite{goerbig_electronic_2011}, a Haldane term~\cite{haldane_model_1988}, or laser-assisted processes~\cite{rudner_band_2020}, there is a different phenomenology evidenced by copropagating hybrid electron-phonon edge states coexisting with a gapless bulk. This new phenomenology stems from the rich interplay between \textit{Umklapp} processes and inelastic effects allowed by the interaction with chiral phonons.

A potential caveat of our study so far is the fact that a single phonon mode is considered. However, the mode is taken into account non-perturbatively and, in the spirit of the Peierls transition, given a mechanism that selects this peculiar high-symmetry phonon mode, the effect of the remaining modes should remain perturbative, thereby lessening their importance~\cite{frohlich_theory_1954,pule_peierls-frohlich_1994}. Furthermore, experiments may specifically target chiral phonons by enhancing the electron-phonon interaction through optical pumping of the selected mode~\cite{gambetta_real-time_2006,kim_coherent_2013}. As for the possibilities of bringing these results to experiments, a possible playground is graphene on hBN~\cite{gao_nondegenerate_2018}. In this case, the substrate provides the inversion symmetry breaking needed to lift the degeneracy between chiral phonons~\cite{gao_nondegenerate_2018}, while providing for an electronic gap around the Dirac point ($\sim 40$ meV) which is smaller than $\hbar\omega$ ($\sim 160$ meV). Our simulations for this case (see ~\cite{supp-info}) show that e-ph interaction strengths of $\gamma_1 = 0.0002\gamma_0 = 5.4 \, \text{meV}$ and $\gamma_1 = 0.002\gamma_0 = 0.54 \, \text{meV}$ give a pseudogap, bridged by copropagating states, in the range of $4.5-45$ meV, accessible through low to moderate cryogenic temperatures. Transport experiments could be aided by scanning gate microscopy~\cite{moreau_upstream_2021}. We hope that our results could foster new experiments in this promising \textit{terra incognita}.

\begin{acknowledgments}
We thank the support of FondeCyT (Chile) under grant number 1211038, and by the EU Horizon 2020 research and innovation program under the Marie-Sklodowska-Curie Grant Agreement No. 873028 (HYDROTRONICS Project). L. E. F. F. T. also acknowledges the support of The Abdus Salam International Centre for Theoretical Physics and the Simons Foundation. H. L. C. is member of CONICET and acknowledges financial support by SECYT-UNC and ANPCyT (PICT-2018-03587). J.M.D. is supported by CONICYT Grant CONICYT-PFCHA/Mag\'{\i}sterNacional/2019-22200526
\end{acknowledgments}


%


\appendix

\vspace{1.5cm}
\textbf{ Supplemental material: Copropagating edge states produced by the interaction between electrons and chiral phonons in two-dimensional materials}
\section{The interaction Hamiltonian: Detailed calculation and zone-folding scheme}
In this section we describe in further detail the Hamiltonian of the system.

First, the passage from Eq.~(3) to Eq.~(4) of the main text is done by imposing  phonon quantization, replacing the harmonic time dependence by bosonic phonon operators, $e^{-i\omega t} \rightarrow a^\dag$ and $e^{i\omega t} \rightarrow a$. The interaction Hamiltonian thus reads
\begin{equation}
\begin{split}
    \mathcal{H}_\text{e-ph} = -\gamma_1 a^\dag \Big[ & \sum_{\left\langle n,m \right\rangle} c_n^\dag c_m e^{i\mathbf{G}\cdot\mathbf{R}_n}\hat{\mathbf{r}}_{n,m} \cdot \\ & \left( \mathbf{u}_\nu - e^{-i\mathbf{G}\cdot\mathbf{R}_{n,m}} \mathbf{u}_\mu \right) \Big] + \text{h.c.} \text{ ,}
\end{split}
\end{equation}
with $\mathbf{R}_{n,m} = \mathbf{R}_n - \mathbf{R}_m$. The e-ph hoppings contain a spatial dependence due to momentum interchange between both particles. Defining the Fourier transform 
\begin{equation}
    c_n = \int_\text{BZ} \text{d}^2\mathbf{k} \, e^{i\mathbf{k}\cdot\mathbf{R}_n} c_{\nu,\mathbf{k}} \text{ ,}
    \label{eq:fouriertransform}
\end{equation}
one gets to Eq.~(6) of the main text:
\begin{equation}
\begin{split}
    \mathcal{H}_\text{e-ph} = & -\gamma_1 a^\dag \Big[ \int_\text{BZ} \text{d}^2\mathbf{k} \sum_{\nu\neq\mu} c_{\nu,\mathbf{k}}^\dag c_{\mu,\mathbf{k}+\mathbf{G}} \sum_m e^{-i(\mathbf{k}+\mathbf{G})\cdot\mathbf{R}_{n,m}} \\ & \hat{\mathbf{r}}_{n,m} \cdot \left( \mathbf{u}_\nu - e^{-i\mathbf{G}\mathbf\mathbf{R}_{n,m}} \mathbf{u}_\mu \right) \Big] + \text{h.c.} \text{ ,}
    \label{eq:e-phhamiltonian}
\end{split}
\end{equation}
where the first sum runs over the sublattices, and the second sum runs over all sites $m$ of sublattice $\mu$, which are nearest neighbors of a given site $n$ of sublattice $\nu$. 

Now, starting from this equation, we define
\begin{equation}
    t_\text{e-ph}(\mathbf{k}) = -\gamma_1 \sum_m e^{-i\mathbf{k}\cdot\mathbf{R}_{n,m}} \hat{\mathbf{r}}_{n,m} \cdot \left( \mathbf{u}_\text{A} - e^{-i\mathbf{G}\cdot\mathbf{R}_{n,m}} \mathbf{u}_\text{B} \right) \text{ ,}
\end{equation}
where the sum runs over all sites $m$ of sublattice B which are nearest neighbor of a given site $n$ of sublattice A. We note as well that the inverse term where site $m$ is of sublattice A and site $n$ is of sublattice B is equal to $t_\text{e-ph}(-\mathbf{k}-\mathbf{G})$. Defining $\psi_\mathbf{k} = (c_{\text{A}, \mathbf{k}}, c_{\text{B}, \mathbf{k}})^\text{T}$, we obtain a matrix representation of the interaction Hamiltonian $\mathcal{H}_\text{e-ph} = a^\dag\int_\text{BZ} \text{d}^2\mathbf{k} \psi_\mathbf{k}^\dag \text{h}_\text{e-ph}(\mathbf{k}) \psi_{\mathbf{k}+\mathbf{G}} + \text{h.c.}$, with
\begin{equation}
    \text{h}_\text{e-ph}(\mathbf{k}) = \begin{pmatrix} 0 & t_\text{e-ph}(-\mathbf{k}-\mathbf{G}) \\ t_\text{e-ph}(\mathbf{k}) & 0 \end{pmatrix} \text{ .}
\end{equation}
The Hamiltonian considers non-vertical electronic transitions in the full BZ, however, we may develop a zone folding scheme where the system presents only vertical transitions in a reduced BZ (rBZ) as follows: We partition the BZ into zones A, B and C as depicted in Fig.2, corresponding to the Voronoi decomposition with respect to points $\Gamma$, $\mathbf{G} = K^{+}$ and $-\mathbf{G} = K^{-}$, where all Voronoi cells present the same geometry. We fold zones B and C into zone A, which corresponds     to the rBZ, expressing the BZ zone integration as
\begin{equation}
\begin{split}
    \int_\text{BZ} \text{d}^2\mathbf{k} f(\mathbf{k}) &= \left( \int_\text{A} \text{d}^2\mathbf{k} + \int_\text{B} \text{d}^2\mathbf{k} + \int_\text{C} \text{d}^2\mathbf{k} \right) f(\mathbf{k}) \\ &= \int_\text{rBZ} \text{d}^2\mathbf{k} \, \left[ f(\mathbf{k}) + f(\mathbf{k} + \mathbf{G}) + f(\mathbf{k} - \mathbf{G}) \right] \text{ .}
\end{split}
\end{equation}
We now define $\Psi_\mathbf{k} = (\psi_\mathbf{k}^\text{T}, \psi_{\mathbf{k}+\mathbf{G}}^\text{T}, \psi_{\mathbf{k}-\mathbf{G}}^\text{T})^\text{T}$, obtaining a matrix Hamiltonian with vertical transitions in the rBZ, $\mathcal{H}_\text{e-ph} = a^\dag \int_\text{rBZ} \text{d}^2\mathbf{k} \Psi_\mathbf{k}^\dag \text{H}_\text{e-ph}(\mathbf{k}) \Psi_\mathbf{k} + \text{h.c.}$, with
\begin{equation}
    \text{H}_\text{e-ph} = \begin{pmatrix} 0 & \text{h}_\text{e-ph}(\mathbf{k}) & 0 \\ 0 & 0 & \text{h}_\text{e-ph}(\mathbf{k}+\mathbf{G}) \\ \text{h}_\text{e-ph}(\mathbf{k}-\mathbf{G}) & 0 & 0 \end{pmatrix} \text{ .}
\end{equation}

In this basis the electronic Hamiltonian is $\mathcal{H}_\text{e} = \int_\text{rBZ} \text{d}^2\mathbf{k} \Psi_\mathbf{k}^\dag \text{H}_\text{e}(\mathbf{k}) \Psi_\mathbf{k}$, with
\begin{equation}
    \text{H}_\text{e}(\mathbf{k}) = \begin{pmatrix} \text{h}_\text{e}(\mathbf{k}) & 0 & 0 \\ 0 & \text{h}_\text{e}(\mathbf{k}+\mathbf{G}) & 0 \\ 0 & 0 & \text{h}_\text{e}(\mathbf{k}-\mathbf{G}) \end{pmatrix} \text{ ,}
\end{equation}
\begin{equation}
    \text{h}_\text{e}(\mathbf{k}) = \begin{pmatrix} \Delta & t_\text{e}^*(\mathbf{k}) \\ t_\text{e}(\mathbf{k}) & -\Delta \end{pmatrix} \text{ ,}
\end{equation}
and $t_\text{e}(\mathbf{k}) = \gamma_0 \sum_m e^{-i\mathbf{k}\cdot\mathbf{R}_{n,m}}$, where the sum runs over all sites $m$ of sublattice B which are nearest neighbor of a given site $n$ of sublattice A. The phonon Hamiltonian is proportional to the identity in the electronic subspace.

We represent the phonon subspace in its Fock basis, obtaining a matrix representation for the full Hamiltonian
\begin{equation}
    \text{H} = \begin{pmatrix} 
    \text{H}_\text{e} & \text{H}_\text{e-ph}^\dag & 0 & \\
    \text{H}_\text{e-ph} & \text{H}_\text{e} + \hbar\omega & \sqrt{2} \text{H}_\text{e-ph}^\dag & \\
    0 & \sqrt{2} \text{H}_\text{e-ph} & \text{H}_\text{e} + 2\hbar\omega & \\ 
     & & & \ddots
    \end{pmatrix} \text{ .}
    \label{eq:fullhamiltonian}
\end{equation}
The diagonal terms in Eq.~\eqref{eq:fullhamiltonian} constitute a semi-infinite series of pure electronic Hamiltonian replicas, centered at energies $n_\text{ph}\hbar\omega$. The off-diagonal elements couple the different replicas.

In the following we proceed to obtain a low-energy approximation. The replica crossing at energy $\hbar\omega/2$ may be described by truncating the full Hamiltonian to both valleys in replicas $n_\text{ph}=0$ and $n_\text{ph}=1$, where only valleys $K^-$ and $K^+$ in replicas $0$ and $1$ respectively are coupled by the e-ph interaction. The coupling is described by the effective Hamiltonian
\begin{equation}
    \text{H}_\text{eff} = \begin{pmatrix}
    \text{h}_\text{e}(\mathbf{k}-\mathbf{G}) & \text{h}_\text{e-ph}^\dag(\mathbf{k}+\mathbf{G}) \\
    \text{h}_\text{e-ph}(\mathbf{k}+\mathbf{G}) & \text{h}_\text{e}(\mathbf{k}+\mathbf{G}) + \hbar\omega
    \end{pmatrix} \text{ .}
\end{equation}
For $\hbar\omega \ll \gamma_0$ we approximate the Hamiltonian about the $K^\pm$ points, obtaining
\begin{subequations}
\begin{equation}
    \text{h}_\text{e}(\mathbf{k}\pm\mathbf{G}) \approx \begin{pmatrix}
    \Delta & \mp \hbar v_\text{F}k e^{\pm i\theta} \\
    \mp \hbar v_\text{F}k e^{\mp i\theta} & -\Delta 
    \end{pmatrix} \text{ ,}
\end{equation}
\begin{equation}
    \text{h}_\text{e-ph}(\mathbf{k}+\mathbf{G}) \approx \begin{pmatrix}
    0 & t_1 - i\mathbf{k}\cdot\mathbf{t}_2 \\
    t_1 + i \mathbf{k}\cdot\mathbf{t}_2 & 0
    \end{pmatrix}
\end{equation}
\end{subequations}
with $\hbar v_\text{F} = 3a_0\gamma_0/2$, $\mathbf{k} = k(\cos\theta \hat{\mathbf{x}} + \sin\theta \hat{\mathbf{y}})$, $t_1 = t_\text{e-ph}(\mathbf{G})$, and $\mathbf{t}_2 = -\gamma_1 \sum_m e^{-i\mathbf{G}\cdot\mathbf{R}_{n,m}} [ \hat{\mathbf{r}}_{n,m} \cdot (\mathbf{u}_\text{A} - e ^{-i\mathbf{G}\cdot\mathbf{R}_{n,m}} \mathbf{u}_\text{B}) ] \mathbf{R}_{n,m}$, where the sum runs over all sites $m$ of sublattice B which are nearest neighbor of a given site $n$ of sublattice A. 

We rotate the basis of the Hamiltonian matrix representation to the one that diagonalizes the blocks $\text{h}_\text{e}(\mathbf{k}\pm\mathbf{G})$. The rotated Hamiltonian is $\tilde{H}_\text{eff} = \text{U}^\dag \text{H}_\text{eff} \text{U}$, with $\text{U} = \text{diag}(\text{U}_-, \text{U}_+)$, and $\text{U}_\pm$ the transformation that diagonalizes $\text{h}_\text{e}(\mathbf{k}\pm\mathbf{G})$. We truncate the space to the upper band of $n_\text{ph}=0$ and the lower band of $n_\text{ph}=1$, obtaining an effective two band model for the avoided crossing
\begin{equation}
    \tilde{\text{h}}_\text{eff} = \begin{pmatrix}
    \sqrt{\Delta^2 + (\hbar v_\text{F}k)^2} & -e^{i\theta} (t_1^* - i\mathbf{k}\cdot\mathbf{t}_2^*) \\
    -e^{-i\theta}(t_1 + i\mathbf{k}\cdot\mathbf{t}_2) & \hbar\omega - \sqrt{\Delta^2 + (\hbar v_\text{F}k)^2}
    \end{pmatrix} \text{ .}
\end{equation}
Within a first approximation we may disregard the term $\mathbf{k}\cdot \mathbf{t}_2$, concluding that the coupling is uniquely determined by $t_1 = -3i[\mathbf{L}^\dag \mathbf{u}_\text{A} + \mathbf{R}^\dag \mathbf{u}_\text{B}] / \sqrt{2}$. The magnitude of the gap is $2|t_1|$, which occurs for $(\hbar v_\text{F}k)^2 = (\hbar\omega/2)^2 - \Delta^2$, and depends on the particular circular motion of each sublattice. For $\mathbf{u}_\text{A} = \mathbf{L}$ and $\mathbf{u}_\text{B} = \mathbf{R}$ we obtain $\mathbf{t}_2 = t_1 \hat{\mathbf{y}}$, which generates a dependence of the avoided crossing on the direction of the momentum. The magnitude of the gap continues to be $2|t_1|$, which now occurs at $\theta=0$.

\section{Klein and armchair ribbons}
\begin{figure}[hbt!]
\centering
\includegraphics[width=\linewidth]{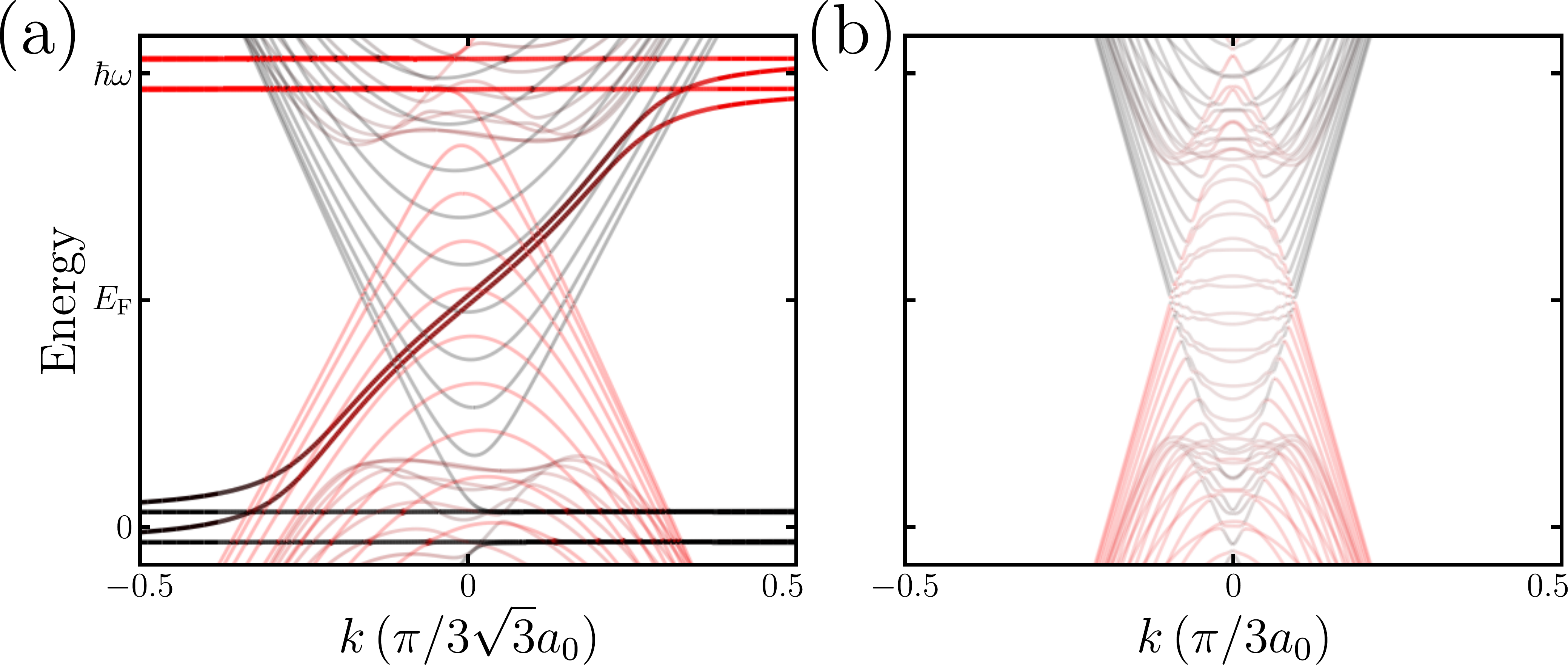}
\caption{\textbf{(a)} Band structure of a Klein ribbon (of width $149a_0$), showing two edge states with propagation direction contrary to those of a zigzag ribbon. \textbf{(b)} Band structure of an armchair ribbon (of width $100\sqrt{3}a_0$). No edge states exist at the replica crossing. Color scale indicates the weight of the bands on $n_\text{ph}=0$ (black) and $n_\text{ph}=1$ (red), while the curves' transparency is related to the localization of the states (see Fig. 3 in the main text).}
\label{fig:fig5}
\end{figure}

In this section we include the band structure for a Klein and an armchair ribbon, shown in Fig.~\ref{fig:fig5}-(a) and (b) respectively. The Klein ribbon presents two copropagating edge states that coexist with a continuum of extended states. However, the Klein states propagate in a direction contrary to those of the zigzag ribbon. In contrast, the armchair ribbon presents no edge states at the replica crossing.

\section{Transport calculations}
The transport calculations presented in Fig.~4 of the main text are aimed at probing transport throughout the ribbon in the non-invasive limit (as in the case of an STM probe for example). This is simulated by placing local probes weakly attached to specific sites of the sample, say $s_1$ located along the dashed line marked with L in the insets of Fig.4 and $s_2$ located along the dashed line marked with R.

These non-invasive probes are modeled through a broadband approximation for the self-energy and the transmission probabilities between these sites, from L to R and vice versa, including elastic and phonon-assisted processes, can be calculated by using the technique in Refs.~\cite{anda_electron_1994,bonca_effect_1995}, that was used for carbon nanotubes in Refs.~\cite{foa_torres_inelastic_2006,foa_torres_nonequilibrium_2008}. Within this picture, the total transmission probability between $s_1$ and $s_2$ is given by:
\begin{equation}
     {\cal T}_{s_2 \leftarrow s_1} = \sum_n 2 \Gamma | G_{(s_2,n),(s_1,0)}^{(r)} |^2 2\Gamma,
\label{eq:ts}
\end{equation}
where $\Gamma$ is the imaginary part of the virtual probe self-energy (which are assumed to have vanishing real part as usual within the broadband approximation) and $G_{(s_2,n),(s_1,0)}^{(r)}$ is the retarded Green's function connecting the orbital at site $s_1$ with $n$ phonons and the orbital at $s_2$ with $n$ phonons. The equilibrium phonon population is assumed to be zero given the large ratio of the phonon energy to relevant temperatures.

\begin{figure}[hbt!]
\centering
\includegraphics[width=\linewidth]{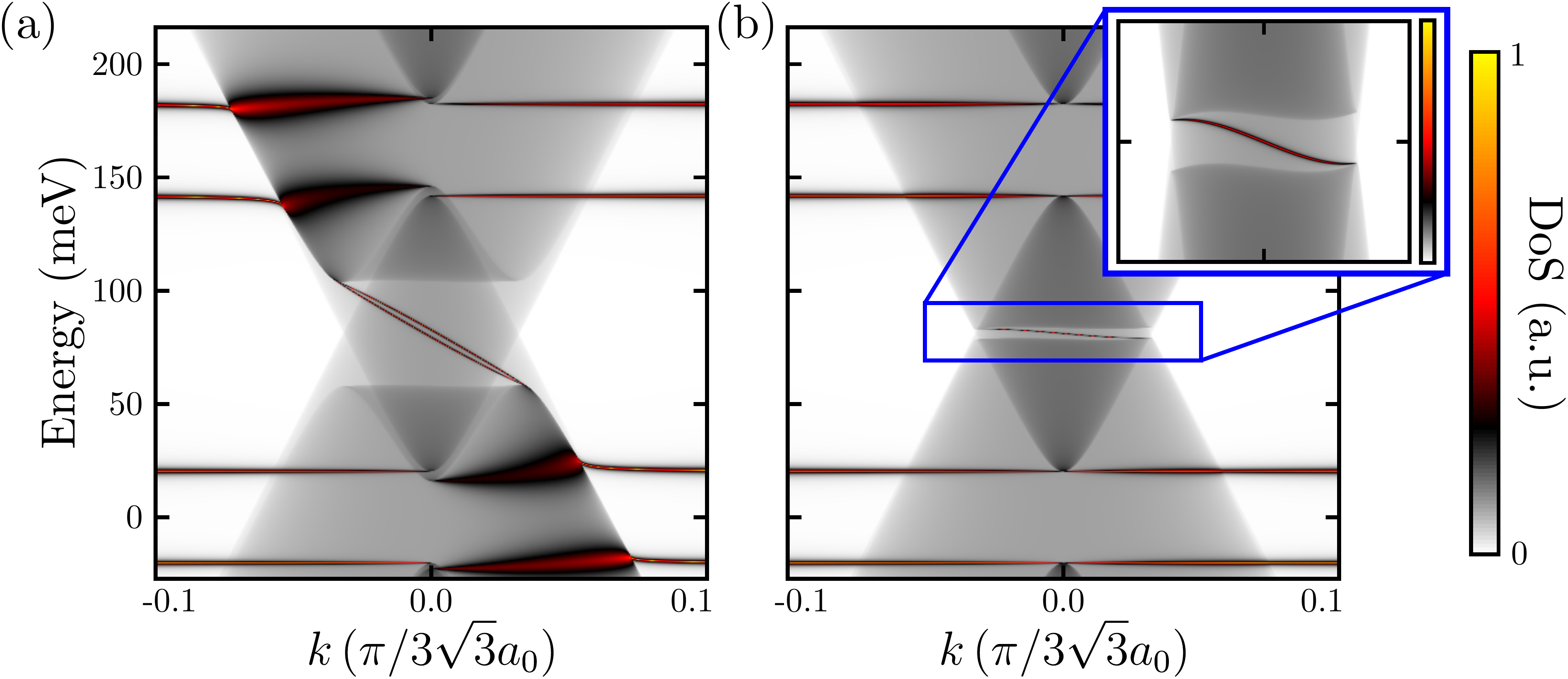}
\caption{Density of States (DoS), shown in logarithmic scale $\text{ln}(1+\text{DoS})$, near the edges of a semi-infinite graphene on hBN zigzag ribbon, in arbitrary units. We use parameter values targeting the case of graphene on hBN; specifically $\gamma_0=2.7\, \text{eV}$, $\hbar\omega=162\, \text{meV}$, and $\Delta=20.25\, \text{meV}$. The e-ph interaction strength is $\gamma_1 = 0.002\gamma_0 = 5.4\, \text{meV}$ in panel (a), and $\gamma_1 = 0.0002\gamma_0 = 0.54\, \text{meV}$ in panel (b), showing that the physics predicted in the main text holds for a wide range of values. 
}
\label{fig:fig6}
\end{figure}

While for invasive contacts one normally uses the same scale to plot transmission probabilities, in the case of non-invasive probes as used in Fig. 4 of the main text this is much less useful. Indeed, in this regime where the contacts are weakly coupled to the sample the transmission probabilities are usually very small (with a probability scaling with the square modulus of the matrix element coupling the contact with the sample). Furthermore, since the probes are floating at different positions, having a reduced transmission probability does not mean that the wave does not propagate in that direction as it might be just spread in space. This is the reason why we chose a normalization that reflects on the space distribution of the probability rather than its magnitude and show that this space distribution (and its directionality) is robust to disorder. For reference the sum of the transmission probabilities (Fig. 4 of the main text) with input site on the right (left) for the disordered system is $0.46$ ($0.48$) that of the pristine one. However the directionality of the transmission and its spatial distribution is little disturbed.

\section{Simulations for graphene on hBN}
In this section we provide more details on a case that may help the experimental search of the physics explained in the main text. Specifically, we put forward the case of graphene on aligned hBN. The presence of carefully aligned substrate allows for the inversion symmetry breaking required to lift the degeneracy between chiral phonons, as shown by recent calculations~\cite{gao_nondegenerate_2018}. But since such a breaking typically produces a bandgap on the electronic structure, one needs to prevent for the gap to be larger than $\hbar\omega$. Otherwise, both valleys would become gapped at the energy range of interest and one would not be able to evidence the physics predicted in the text (a valley-selective pseudogap bridged by copropagating edge states). Fortunately, for graphene on hBN this is usually the case, as the gap is on the order of about $30-40$ meV (see ~\cite{ribeiro-palau_twistable_2018}) while $\hbar\omega \sim 160$meV (which, in turn, is much smaller than $\gamma_0 \simeq 2.7$eV).

Furthermore, placing the Fermi level at the pseudogap requires a gating of about $80$ meV ($\hbar\omega/2$) which is feasible with current techniques (which allow up to several hundreds of meV of effective gating). Fig.~\ref{fig:fig6} shows a color plot of the the density of states for the parameters estimated for graphene on hBN and two values of the e-ph coupling strength. The resulting pseudogaps are of (a) $45.4$ meV and (b) $4.5$ meV which are in good agreement with the analytical estimations ($45.8$ meV and $4.6$ meV, respectively). This would require low to moderate cryogenic temperatures ($\sim 10-100$ K).

\end{document}